\documentclass[11pt,a4paper]{article}
\usepackage{jheppub}

\usepackage{amsmath,graphicx,amsfonts,mathrsfs,amssymb}
\usepackage{amsmath,epsfig}
\usepackage{amssymb,amsfonts}
\usepackage{latexsym}
\usepackage{epsfig}
\newbox\pippobox
\def\be{\begin{equation}}
\def\ee{\end{equation}}
\def\bea{\begin{eqnarray}}
\def\eea{\end{eqnarray}}

\def\ee           {{\rm e}}

\newcommand{\beq}{\begin{equation}}
\newcommand{\eeq}{\end{equation}}
\newcommand{\beqa}{\begin{eqnarray}}
\newcommand{\eeqa}{\end{eqnarray}}
\newcommand{\beqar}{\begin{eqnarray*}}
\newcommand{\eeqar}{\end{eqnarray*}}

\renewcommand{\eqref}[1]{(\ref{#1})}

\catcode`\@=12

\title{Note on Entanglement Temperature for Low Thermal Excited States in Higher Derivative Gravity}
\author[a,b,c]{Wu-zhong Guo,}
\author[a,b]{Song He,}
\author[d,a,b]{Jun Tao}

\affiliation[a]{State Key Laboratory of Theoretical Physics,
Institute of Theoretical Physics, Chinese Academy of Science,
Beijing 100190, People's Republic of China } \affiliation[b]{ Kavli
Institute for Theoretical Physics China, CAS, Beijing 100190, China} \affiliation[c]{University of the Chinese Academy of Sciences, Beijing 100049, China}
\affiliation[d]{Center for Theoretical Physics, College of Physical
Science and Technology, Sichuan University, Chengdu, 610064, PR
China}

\emailAdd{wuzhong@itp.ac.cn}\emailAdd{hesong@itp.ac.cn}\emailAdd{taojun@scu.edu.cn}

\date{\today}

\abstract{We investigate the entanglement temperature of a small
scale subsystem in low excited states by using holographic method.
Especially, we study the entanglement entropy and entanglement
temperature in higher derivative gravities which are considered as
low thermal excitation of pure AdS gravity. We find that the
entanglement entropy are related to the central charges of CFT living on the boundary.
 The relation between the variance of entanglement entropy and
energy of a small scale subsystem has been also obtained.
Furthermore, the relation is consistent with the first law-like
relation that is proposed by Phys. Rev. Lett. 110, 091602 (2013).
Finally, we derive the formula of the variance of entanglement
entropy in general excited states in gravity background with the
Fefferman-Graham coordinates and the entanglement temperature can be
figured out in special case.}


\begin{document}

\maketitle

\section{Introduction}
The AdS/CFT correspondence
\cite{Maldacena:1997re}\cite{Gubser:1998bc}\cite{Witten:1998qj}\cite{Aharony:1999ti}(also
called by gauge/gravity duality) is a very important and fundamental
relation which connects gravitational theories and quantum field
theories. As an application in \cite{Ryu:2006bv} Ryu and Takayanagi
proposed a frame work for calculation of entanglement entropy of
conformal field theory through AdS/CFT correspondence. Their approach
is simple and elegant. The main point is that the entanglement
entropy in field theory side can be mapped to an area of minimal
surface in gravity side. The holographic entanglement entropy (HEE)
have been proved in \cite{Lewkowycz:2013nqa}\cite{Casini:2011kv}.
And there are so many evidences
\cite{Headrick:2010zt}\cite{Hartman:2013mia}\cite{Faulkner:2013yia}
to confirm the correspondence within $AdS_3/CFT_2$. As application
of HEE, there are intensive studies
\cite{Nishioka:2009un,Albash:2011nq,Myers:2012ed,deBoer:2011wk,Hung:2011xb,Nishioka:2006gr,Sun:2008uf,
Klebanov:2007ws,Pakman:2008ui,Ogawa:2011fw,Cai:2012sk,Cai:2012nm,Cai:2012es}
recently. More recently, in \cite{Nozaki:2013wia}, a free falling
particle in an AdS space mimic the holographic dual of local
quenches and the HEE has been computed to show the evolution of
quantum entanglement. In \cite{Hartman:2013qma}, an analytical
framework for holographic counterpart of global quantum quenches. In
\cite{Nozaki:2013vta}, authors studied how a small perturbation of
HEE evolves dynamically through solving the Einstein equation in AdS
spaces.\\
In vacuum state entanglement entropy(EE) is proportional to the
surface area(in many models)\cite{MBH}\cite{Srednicki:1993im}
 in the leading divergent term, which is the original motivation for relating EE with black hole entropy. The EE is
also an useful quantity to describe the quantum correlations between
the in and out side of a subsystem in QFT. The behavior of EE in low excited states is also
important to understand the quantum entanglement nature of the
system. This topic has been studied by many authors, for example
\cite{FMG}\cite{Masanes:2009tg}. The elegant method of HEE could also be
used to study the property of EE in low excited states of CFT, which
may be related with the background perturbation of the bulk.

In \cite{Bhattacharya:2012mi} the authors go on this study about the
low thermal excited state in the holographic view, and more, they
find an interesting relation between the variance of energy and EE
for the subsystem in low excited states of CFT, which is similar as
the first law of thermodynamics, i.e., $\Delta E =T_{eff} \Delta S$,
where they also introduce a universal quantity $T_{eff}$ called
entanglement temperature, which is only related with shape of the
subsystem.

In this paper we will study the property of EE after low excitation
in the holographic view. Considering that the HEE formula should
have quantum correction when the bulk theory has higher curvature
terms\cite{deBoer:2011wk}\cite{Hung:2011xb}\cite{Fursaev:2006ih}, we
expect new property of EE would appear in low excited states. The
general formula of HEE with the bulk theory containing arbitrary
higher curvature terms is still a open question needed to further
study. But the HEE formula for Lovelock gravity has been studied in
\cite{deBoer:2011wk}\cite{Hung:2011xb} by comparing the logarithm
term with the CFT prediction. In terms of
\cite{Hung:2011xb}\cite{deBoer:2011wk}\cite{Ogawa:2011fw}\cite{Myers:2010xs},
one can study the HEE with higher derivative gravity and see what
will happen for the variance of EE of a small scale subsystem.
Furthermore, the energy momentum tensor will also include
contribution from higher derivative gravity. So it would also be
interesting to investigate the first law-like relation and
entanglement temperature that are proposed in
\cite{Bhattacharya:2012mi} then.

This paper is organized as follows. In section 2 we calculate the
entanglement temperature with roll ball profile and infinite stripe
profile with a small scale in the 5-dimensional Lovelock gravity and
study correction of the first law-like relation for these
subsystems. In section 3, we extend to the holographic entanglement
entropy and entanglement temperature for 7-dimensional Lovelock
gravity. In section 4, we analyze the holographic entanglement
entropy and entanglement temperature in asymptotical AdS with
Fefferman-Graham gauged background in diverse dimension. Section 5
is devoted to conclusions and discussions. We put some details of
the computations in this paper in the Appendix.

While proceeding with this project, we learned that the small
overlap topic in section 4 was also studied in recent papers
\cite{Nozaki:2013vta}\cite{Allahbakhshi:2013rda} which are reminded by
Tadashi Takayanagi and Mohsen Alishahiha.

\section{Entanglement temperature in D=4 CFT with dual 5-dimensional Lovelock gravity}
In this section, we will calculate the entanglement temperature in
the 4-dimensional CFT with the dual 5-dimensional Lovelock gravity.
To obtain the novel quantities, we should calculate the variance of
entanglement entropy after low thermal excitation in two examples,
i.e., the entangling surfaces are sphere and infinite stripe, and
also the variance of the energy within the subsystem.
\subsection{HEE for 4-dimensional CFT}
The 5-dimensional Lovelock gravity can be realized by adding the
Gauss-Bonnet term\cite{Lovelock} to pure Einstein gravity theory.
The theory can be described by the following action
\begin{eqnarray}
I = \frac{1}{2{\ell_p}^3} \int \mathrm{d}^5x \, \sqrt{-g}\, \left[ R+
\frac{12}{L^2}  + \frac{\lambda_5 L^2}{2} L_4 \right],
 \end{eqnarray}
with
\begin{eqnarray}
 L_4=R_{\mu\nu\rho\sigma}R^{\mu\nu\rho\sigma}-4R_{\mu\nu}R^{\mu\nu}+R^2,
\end{eqnarray} and $\lambda_5$ denotes the coupling
of Gauss-Bonnet gravity and $L$ stands for the Radius of AdS
background. In order to obtain a well-defined vacuum for the gravity
theory, one has to have $\lambda_5 \leq 1/4$. Additional constraint
on the Gauss-Bonnet parameter $\lambda_5$ by considering the
positivity of the energy flux and causality on the boundary theory
is $-7/36\leq \lambda_5 \leq
9/100$~\cite{brigante07,Brigante:2008gz,Buchel:2009tt,Hofman:2009ug}.
The AdS Gauss-Bonnet (GB) gravity would admit a pure AdS
solution\cite{Hung:2011xb},
\begin{eqnarray}
 ds^{2}=\frac{\tilde
{L}^2}{z^2}(-dt^2+dz^2+dx_1^2+dx_2^2+dx_3^2)
\label{PADS},
\end{eqnarray}
$\tilde {L}$ is the effective AdS radius in Gauss-Bonnet gravity and
is defined by $\tilde {L}^2$=$\frac{L^2}{f_{\infty}}$ with
\begin{eqnarray}\label{finfinity}
f_{\infty}=\frac{1-\sqrt{1-4\lambda_5}}{2\lambda_5}.
\end{eqnarray} The holographic entanglement entropy formula for
the GB case is discussed in  \cite{deBoer:2011wk}\cite{Hung:2011xb}
\cite{Fursaev:2006ih}. The subsystem in a time slice of boundary is
$A$, we recall that the HEE formula should be
\begin{eqnarray}
S_A = \frac{2 \pi}{\ell_p^3} \int_M d^3x\sqrt{h}\,\left[ 1+ \lambda_5
L^{2} \mathcal{R} \right]+\frac{4\pi}{\ell_p^{3}} \int_{\partial M}
d^{2}x\sqrt{h} \lambda_5 L^{2} \mathcal{K}, \label{GBEE4}
\end{eqnarray}
where the first integral is evaluated on the bulk surface $M$, the second one is on $\partial M$, which is the boundary of M regularized at $z=\epsilon$, $\mathcal{R}$ is the Ricci scalar for the intrinsic
geometry of $M$, and $\mathcal{K}$ is the trace of the extrinsic
curvature of the boundary of $M$, $h$ is the determinant of the
induced metric on $M$. The second term in the first integral is
present due to higher derivative gravity in the background. The
minimal value of the functional ($\ref{GBEE4}$) would give the
entanglement entropy of the subsystem $A$.
\subsection{Variance of EE after low excitation}
Our purpose is to discuss the variance of the entanglement entropy
for the low excited state of CFT from the holographic point of view.
Following the logic given by \cite{Bhattacharya:2012mi}, the low
exciting state of the CFT is dual to the asymptotical AdS background
and the pure AdS solution can be considered as the ground state of
AdS GB gravity. Here we will assume the low excited state with the
dual gravity background as\footnote{We recall that the 5-dimensional
Gauss-Bonnet-AdS black brane solution with a Ricci flat horizon is
given by~\cite{Cai}
\begin{eqnarray}
ds^2_{BB}
=\frac{L^2}{z^2}\Big[-f(z)dt^2+dx_1^2+dx_2^2+dx_3^2+\frac{dz^2}{f(z)}\Big],\nonumber
\\
f(z) =\frac{1}{2\lambda_5}\bigg(1-\sqrt{1-4\lambda_5\big(
1-\frac{z^4}{z_h^4}\big)}\bigg), \nonumber
\end{eqnarray}
where $z_h$ is the horizon of the black brane. The asymptotical AdS
background ($\ref{AADS}$) is actually the above solution with the
horizon $z_h$ far away from the boundary. So we may call this kind
of excited state of the CFT living on the boundary
as the low thermal excitation.}\cite{Bhattacharya:2012mi}
\begin{eqnarray}\label{peturbation}
ds^{2}=\frac{\tilde
{L}^2}{z^2}(-f(z)dt^2+\frac{dz^2}{g(z)}+dx_1^2+dx_2^2+dx_3^2),
\label{AADS}
\end{eqnarray}
where $f(z)\simeq g(z)=1-mz^4$, $m$ reflects the asymptotical behavior of gravity background near the boundary. We identify the
excitation as low thermal excitation in this paper.

In the next two subsections, we would like to discuss two special
examples to see the behavior of the variance of the entanglement
entropy after the low thermal excitation ($\ref{AADS}$).

\subsubsection{Subsystem with a round ball configuration}
In this subsection, we choose the subsystem $A$ as a two-sphere with
a radius $R_{0}$ in a time slice of the CFT living
on the boundary. The bulk surface $M$ can be parameterized as
$r=r(u)$ and $z=z(u)$, where $r$ is the radial direction of the
ball, with $u_{min} \le u \le u_{max}$. So the induced line element
on the bulk surface is
\begin{eqnarray}
ds^2_D=\frac{\tilde {L}^2}{z^2}\Big[\Big((1+mz^4)\dot {z}^2+\dot
{r}^2\Big)du^2+r^2d\Omega_{2}^2\Big],
\end{eqnarray}
where the dot denotes the derivative with respect to
$u$. Following the notation of \cite{Hung:2011xb}, we define
\begin{eqnarray}
h_{uu}={h^{uu}}^{-1}=\frac{\tilde {L}^2}{z^2}\left((1+mz^4)\dot {z}^2+\dot
{r}^2\right), \qquad F=\ln \left(\frac{r}{z}\right).
\end{eqnarray}
We can also get the intrinsic Ricci scalar of $M$
\begin{eqnarray}
\mathcal{R}_D=e^{-2F}\frac{2}{\tilde{L}^2}-4\Delta_u F-6h^{uu}{\dot{F}}^2,
\label {RC}
\end{eqnarray}
where $\Delta_u F=\frac{1}{\sqrt{h_{uu}}}\partial_u(\frac{1}{\sqrt{h_{uu}}}\dot{F})$.
The trace of extrinsic curvature $\mathcal {K}$ of the boundary
$\partial M$ is
\begin{eqnarray}
\mathcal {K}=-\frac{2}{\sqrt{h_{uu}}}\dot {F} |_{u=u_{min}}
\label {EC}
\end{eqnarray}
Plug (\ref{EC}) and ($\ref{RC}$) into ($\ref{GBEE4}$), one can
obtain
\begin{eqnarray}
S_A=\frac{8\pi ^2 \tilde {L}^2}{\ell_p^3}\int du \sqrt{h_{uu}}
e^{2F}(1+L^2 \lambda_5 (e^{-2F}\frac{2}{\tilde{L}^2}+2h^{uu} (\dot
F)^2)).
\end{eqnarray}
It is convenient to parameterize $r$ and $z$ as following form
\begin{eqnarray}
r(u)=f\left({u\over R_0}\right)\cos\left({u\over R_0}\right),\quad
z(u)=f\left({u\over R_0}\right)\sin\left({u\over R_0}\right), \quad
\text{with} \quad \epsilon \le u \le \frac{\pi}{2}R_0.
\end{eqnarray}
Finally, we get $S_A$ as a functional of $f({u\over R_0})$. The
minimal value of functional $S_A$ corresponds to the holographic
entanglement entropy. We first need to get the surface that
minimizes the functional when the perturbation parameter $m$ is turned
off. $S_A$ is functional of $f(x)$ as
\begin{eqnarray}\label{ball-AdS}
S_{A}=\frac {8\pi ^2 \tilde{L}^3}{\ell_p^{3}} \int_{\epsilon
/R_0}^{\frac{\pi}{2}} dx \frac {\cos^2(x)} {\sin^3(x)}
\sqrt{1+(\frac{d \log f}{dx})^2}\left(1+2(\frac{L}{\tilde{L}})^2
\lambda_5 (\tan^2(x)+\frac{\cos^{-2}(x)}{1+(\frac{d \log
f}{dx})^2})\right),\nonumber \\
\end{eqnarray}
where we denote $x=u\over R_0$. One should note that we should introduce the regulation procedure to
make the $S_A$ to be finite. The regulation scale is denoted by the
$\epsilon$ in (\ref{ball-AdS}). Using Euler-Lagrange equation, we
get
\begin{eqnarray}
\frac{d}{dx}\Big[\frac{d \log f}{dx}\frac {\cos^2(x)} {\sin^3(x)} \Big(\frac{1+2f_\infty \lambda_5 \tan^2(x)}{\sqrt{1+(\frac{d \log f}{dx})^2}}-\frac{f_\infty \lambda_5 \cos^{-2}(x)}{(1+(\frac{d \log f}{dx})^2)^{3/2}}\Big) \Big]=0.
\end{eqnarray}
One solution is that $f(x)$ is a constant. We fix the boundary
condition as $f({u\over R_0})=R_0$. The the minimal surface can be
parameterized as
\begin{eqnarray}
 r(u)=R_0 \cos(u/R_0),\quad z(u)=R_0 \sin(u/R_0), \quad \text{with} \quad \epsilon \le u \le \frac{\pi}{2}R_0.
 \label{SP}
 \end{eqnarray}
 The holographic entanglement entropy for pure AdS state is
 \begin{eqnarray}
 S_{A}=\frac {8\pi ^2 \tilde{L}^3}{\ell_p^{3}} \int_{\epsilon /R_0}^{\frac{\pi}{2}} dx \frac {\cos^2(x)} {\sin^3(x)} \Big(1+2(\frac{L}{\tilde{L}})^2 \lambda_5 (\tan^2(x)+\cos^{-2}(x))\Big).
 \end{eqnarray}
We formally turn on  $m \neq 0$ (\ref{peturbation})and consider
it as a perturbation of the pure AdS background. Finally, we can
consider the AdS-Gauss-Bonnet gravity with $m$ as an
excitation state with comparing with pure AdS with radius
$\tilde{L}$.  Then the holographic entanglement entropy after the
excitation is the minimal value of the functional
 \begin{eqnarray}
 \tilde{S}_A=\frac {8\pi ^2 \tilde{L}^3}{\ell_p^{3}} \int_{\epsilon /R_0}^{\frac{\pi}{2}} dx \frac {\cos^2(x)}
 {\sin^3(x)}\sqrt{1+mz^4\dot{z}^2} \Big(1+2(\frac{L}{\tilde{L}})^2 \lambda_5 (\tan^2(x)
 +\frac{\cos^{-2}(x)}{1+mz^4\dot{z}^2})\Big).
 \label{4GBSAT}
 \end{eqnarray}
 We can calculate the variance of the holographic entanglement entropy up to $O(mR_0^4)$ as follows
 \begin{eqnarray}
 \Delta S_A=\tilde{S}_A-S_A
 \end{eqnarray}
 Considering the limit $mR_0^4\ll 1$, we expand $(\ref{4GBSAT})$ to first order of $mR_0^4\ll 1$ and use $(\ref{SP})$ and then the variance of holographic entanglement entropy up to $O(mR_0^4)$ is
 \begin{eqnarray}\label{variationS}
 \Delta S_A &=& \frac {8\pi ^2 \tilde{L}^3}{\ell_p^{3}} m R_0^4\int_{\epsilon /R_0}^{\frac{\pi}{2}} dx (\frac{1}{2}\sin x\cos^4x-\lambda_5 \frac{L^2}{\tilde{L}^2}\sin x\cos^4x)
 \nonumber\\
 && =\frac {8\pi ^2 \tilde{L}^3}{\ell_p^{3}} (\frac{1}{10}-\frac{1}{5}\lambda_5 f_{\infty})m R_0^4.
 \end{eqnarray}
In the procedure of getting (\ref{variationS}), we assume that the
bulk surface that corresponds to the minimal functional does not
change with turning on excitation. As appendix A shows, the
configuration of bulk surface is protected after the low excitation
up to $O(mR_0^4)$. As a consistent check, we reproduce the result given
in \cite{Bhattacharya:2012mi} with the Gauss-Bonnet parameter
$\lambda_5=0$. According to AdS/CFT correspondence, the Gauss-Bonnet
gravity parameters $\lambda_5$ would related with the two kinds of
central charges of the CFT side. The A-type and B-type central
charges \cite{Nojiri:1999mh} \footnote{Here we have used convention
  \begin{eqnarray}
  \langle T^i_{\ i}\rangle=\frac{c}{16\pi^2}C_{ijkl}C^{ijkl}-\frac{a}{16\pi^2}(R_{ijkl}R^{ijkl}-4R_{kl}R^{kl}+R^2)\nonumber.
  \end{eqnarray} Where the $\langle T^i_{\ i}\rangle$ stands for the trace anomaly which have been well studied and $a,c$ stands for
  different kinds of central charges.
  The $a$ and $c$ are equal at the limit $\lambda_5\to 0$ \cite{Hung:2011xb}.}
 \begin{eqnarray}
c= \pi^2\frac{\tilde {L}^3}{l_{p}^3}\left(1-2\lambda_5 f_{\infty}
\right)\ ,
 \qquad
a=\pi^2\frac{\tilde {L}^3}{l_{p}^3}\left(1-6\lambda_5 f_{\infty}
\right)\ . \label{CCAC}
\end{eqnarray}
 According to($~\ref{CCAC}$), we can rewrite $\Delta S_A$ as
 \begin{eqnarray}
{\Delta S_A=\frac {4 c}{5}m R_0^4.}\label {DeltaS}
 \end{eqnarray}
One can see that the variance of the entanglement entropy after the
low thermal excitation is related to the central
charges of CFT. One will also see this property in entanglement
entropy of subsystem with an infinite stripe profile.

\subsubsection{Subsystem with stripe configuration}
In this subsection, we consider the subsystem with a stripe profile
which is defined by $-\frac{l}{2}<x_1\equiv x<\frac{l}{2}$ and
$\frac{-l_0}{2}<x_2,x_3<\frac{l_0}{2}$ where $l_0$ is infinite. We
also consider the AdS Gauss-Bonnet solution (\ref{AADS}) as a
excitation of pure AdS gravity. Then the induced metric $h_{\mu\nu}$
of the bulk surface after perturbation (or excitation) is
\begin{eqnarray}
ds_S^2=\frac{\tilde
{L}^2}{z^2}[(\dot{x}^2+1+mz^4)dz^2+dx_2^2+dx_3^2].
\end{eqnarray}
Where $x$ is parameterized as $x=x(z)$ and the dot stands for the
derivative with respect to $z$ in this subsection. We denote
\begin{eqnarray}
h_{zz}=\frac{\tilde{L}^2}{z^2}(\dot{x}^2+1+mz^4),\quad e^{2F}=\frac{\tilde{L}^2}{z^2}.
\end{eqnarray}
The intrinsic Ricci curvature $\mathcal R_S$ is
\begin{eqnarray}
\mathcal R_S=-\frac{4}{\sqrt{h_{zz}}}\partial _z(\sqrt{h_{zz}}h^{zz}\dot{F})-6h^{zz}\dot{F}^2.
\label {RSS}
\end{eqnarray}
The trace of extrinsic curvature $\mathcal {K}$ of the boundary
$z=\epsilon$ is
\begin{eqnarray}
\mathcal {K}=-\frac{2}{\sqrt{h_{zz}}} \dot {F} |_{z=\epsilon}.
\label{ESS}
\end{eqnarray}
Plug ($\ref{RSS}$) and ($\ref{ESS}$) into ($\ref{GBEE4}$)
\begin{eqnarray}
S_A = \frac{2 \pi l_0^2}{\ell_p^3} \int dz\sqrt{h_{zz}}\frac{\tilde{L}^2}{z^2}(1+2h^{zz}\dot{F}^2).
\end{eqnarray}\\
Firstly let's turn off the perturbation, then the bulk is the pure AdS space. $S_A$ is a functional of $x(z)$ as follows
\begin{eqnarray}
S_A (m=0)= \frac{4 \pi l_0^2}{\ell_p^3}\int_\epsilon^{z_*} dz
\frac{\tilde{L}^3}{z^3}(\sqrt{1+\dot{x}^2}+2\lambda_5
\frac{L^2}{\tilde{L}^2}\frac{1}{\sqrt{1+\dot{x}^2}}), \label{EESF}
\end{eqnarray}
where $z_*$ is the maximal value of $z$ on the surface in the bulk,
which is controlled by the following constraint
\begin{eqnarray}
\frac{l}{2}=\int^{z_*}_0dz\dot{x},
\label{SC}
\end{eqnarray}
with minimizing the functional ($\ref{EESF}$), we get the following
equation of motion
\begin{eqnarray}
\dot{x}\frac{1-2\lambda_5 f_\infty+\dot{x}^2}{(1+\dot{x}^2)^{3/2}}=\frac{z^3}{z_*^3}
\label{SEM}
\end{eqnarray}
This is a cubic equation for $\dot{x}$, there are three solutions for
$\dot{x}$. We are only interested in the solution of this equation
with $\lambda_5$ is near 0. In this case, we can continuously connect
AdS Gauss-Bonnet gravity with the solution of $\lambda_5=0$, i.e., the
pure Einstein gravity. Here we consider the pure AdS gravity as a ground
state of AdS GB gravity. Assuming that $\lambda_5$ is close to 0 and
$\lambda_5 f_\infty \ll 1$, we can solve the equation ($\ref{SEM}$)
with expanding as small parameter $\lambda_5 f_\infty$. Up to the
$O(\lambda_5 f_\infty)$, we get
\begin{eqnarray}
\dot{x}=(1+2\lambda_5 f_\infty)\frac{\frac{z^3}{z_*^3}}{\sqrt{1-\frac{z^6}{z_*^6}}}.
\label{SFX}
\end{eqnarray}
Using the constraint ($\ref{SC}$), we can get  $z_*$
\begin{eqnarray}
z_*=\frac{l\Gamma(\frac{1}{6})}{2(1+2\lambda_5 f_\infty)\Gamma(\frac{2}{3})\sqrt{\pi}}.
\end{eqnarray}
With turning on the excitation (\ref{peturbation}), the functional
of the entangling surface is
\begin{eqnarray}
S_A(m)= \frac{4 \pi l_0^2}{\ell_p^3}\int_\epsilon^{z_*} dz
\frac{\tilde{L}^3}{z^3}(\sqrt{1+mz^4+\dot{x}^2}+2\lambda_5
\frac{L^2}{\tilde{L}^2}\frac{1}{\sqrt{1+mz^4+\dot{x}^2}})
\end{eqnarray}
Using the conclusion in Appendix A we get the variance of the HEE is
\begin{eqnarray}
\Delta S_A &=& S_A(m)-S_A(m=0)\nonumber \\
&=& \frac{2m \tilde{L}^3 \pi l_0^2}{\ell_p^3}\int_\epsilon^{z_*} dz
z \frac{1-2\lambda_5 f_\infty+\dot{x}^2}{(1+\dot{x}^2)^{3/2}},
\label{MSFS}
\end{eqnarray} Use $ml^4 \ll 1$
and plug the solution of $\dot{x}$ ($\ref{SFX}$) into ($\ref{MSFS}$):
\begin{eqnarray}
\Delta S_A &=& \frac{2m \tilde{L}^3 \pi l_0^2z_*^2}{(1+2\lambda_5 f_\infty)\ell_p^3}\int_{\epsilon/z_*}^{1} du u \sqrt{1-u^6}\nonumber \\
&=& \frac{m \tilde{L}^3 \sqrt{\pi} l_0^2 l^2}{20(1+2\lambda_5 f_\infty)^3 \ell_p^3}\frac{\Gamma(\frac{1}{3})\Gamma(\frac{1}{6})^2}{\Gamma(\frac{2}{3})^2 \Gamma(\frac{5}{6})}\nonumber \\
&\simeq & \frac{a}{2\pi^2}\frac{m  \sqrt{\pi} l_0^2
l^2}{10}\frac{\Gamma(\frac{1}{3})\Gamma(\frac{1}{6})^2}{\Gamma(\frac{2}{3})^2
\Gamma(\frac{5}{6})} \label{MSFS},
\end{eqnarray}
where we use ($\ref{CCAC}$) in the last step. $\Delta S$ depends on
the A-type central charge of CFT living on the
boundary with the assumption that $\lambda_5 f_\infty \ll 1$.
(\ref{MSFS}) will reproduce result in \cite{Bhattacharya:2012mi} with
$\lambda_5\rightarrow 0$.

\subsection{Variance of energy and EE}
In \cite{Bhattacharya:2012mi} they have proposed a universal
relation between the variance of the energy and the entanglement
entropy for a small subsystem A on the boundary theory. The
universal relation induce a novel concept called by entanglement
temperature. We would like to obtain the entanglement temperature in
higher derivative gravity. We apply to calculate the boundary
energy-stress tensor in the low excitation AdS Gauss-Bonnet
background in Appendix C. The $t-t$ component of the energy-stress
tensor  \cite{KS}\cite{SKS} corresponds to the energy density
(\ref{ETGB})
\begin{eqnarray}
T_{tt}=\frac{3 m \tilde{L}^3(1-2 \lambda_5 f_\infty)}{2
\ell_p^3}
\end{eqnarray}\\
For the sphere case, the variance of energy in the subsystem is
\bea \Delta E=\frac{2 \pi m(1-2 f_{\infty}\lambda_5)
\tilde{L}^3R_0^3 }{ \ell_p^3} .\eea The
entanglement temperature defined by \cite{Bhattacharya:2012mi} is
 \begin{eqnarray}
 \frac{1}{T_{ent}}=\frac{\Delta S}{\Delta E}=\frac{2\pi}{5}R_0.
 \end{eqnarray}
We find that the relation is same as the result of
\cite{Bhattacharya:2012mi} for the 2-sphere in the 4-dimensional
CFT. The dependence of the central charges of variance of
entanglement entropy do not appear for entanglement temperature. The
entanglement temperature is proportional to the inverse of $R_0$. As we will see in
d=7 dimensional Lovelock gravity the sphere would also admit this
property in the next section.

For the infinite stripe, the variance of energy in the subsystem is

\bea \Delta E=\frac{3 (1-2f_{\infty} \lambda)m l\tilde{L}^3l_0^2 }{
2\ell_p^3} .\eea The entanglement temperature is
\begin{eqnarray}\label{universal-relation-GB-stripe}
\frac{1}{T_{ent}}=\frac{\Delta S}{\Delta
E}=\frac{a}{c}\frac{ \sqrt{\pi}
}{30}\frac{\Gamma(\frac{1}{3})\Gamma(\frac{1}{6})^2}{\Gamma(\frac{2}{3})^2
\Gamma(\frac{5}{6})}l.
\end{eqnarray}
Where we have used ($\ref{CCAC}$). The entanglement temperature in
infinite stripe subsystem is similar to the sphere case, but
entanglement temperature depends on the two kinds of central charges
of CFT living on the boundary.

In limit $ml^4\to 0$, the entanglement entropy in CFTs, as shown in
(\ref{universal-relation-GB-stripe}), satisfies a universal relation
analogous to the first law of thermodynamics proposed by
\cite{Bhattacharya:2012mi}.  Comparing with the exact result given
by \cite{Bhattacharya:2012mi}, one can find that the coefficient of
$l$ have included the contribution from higher derivative gravity.
With taking the $\lambda=0$, one can reproduce the coefficient of
$l$ in \cite{Bhattacharya:2012mi}. We find that the coefficient of $l$
is no longer universal constant that is only related to the shape of
the entangling surface, but may be related with the central charges
of the CFT.

\section{Entanglement temperature in D=6 CFT with dual 7-dimensional Lovelock gravity}
In this section, we will continue to study entanglement temperature
in the D=6 CFT with dual 7-dimensional Lovelock gravity.
\subsection{HEE for 6-dimensional CFT}
The cubic gravity interaction would appear in Lovelock
gravity in 7 dimensional Lovelock gravity\cite{Lovelock}. The 7-dimensional Lovelock gravity action is
\begin{eqnarray}
I=\frac{1}{2\ell_p^5}\int d^{7}x \sqrt{-g}
(\frac{30}{L^2}+R+\frac{L^2}{12}\lambda_7
\mathcal{L}_4(R)-\frac{L^4}{24}\mu_7 \mathcal{L}_6(R)),
\end{eqnarray}
where $\mathcal{L}_4(R)$ is the curvature-squared term (i.e., the
Gauss-Bonnet term) and $\mathcal{L}_6(R)$ is the cubic term, which
is as follows \cite{Hung:2011xb}{\begin{eqnarray}\label{EulerDensity6}
\mathcal{L}_6(R)=\frac{1}{2^3}\delta^{\nu_1 \nu_2...\nu_6}_{\mu_1
\mu_2...\mu_6}R^{\mu_1\mu_2}_{\quad \nu_1
\nu_2}...R^{\mu_5\mu_6}_{\quad \nu_5 \nu_6},
\end{eqnarray}}
where $\delta^{\nu_1 \nu_2...\nu_6}_{\mu_1 \mu_2...\mu_6}$ is the
totally antisymmetric product of Kronecker delta symbols.

The 7 dimensional Lovelock gravity also admit the pure AdS solution
with the effective radius $\tilde{L}^2=\frac{L^2}{f_\infty}$,
$f_\infty$ is one special root of following equation
\begin{eqnarray}\label{INFITY7}
1=f_\infty-f_\infty^2 \lambda_7 -f_\infty^3\mu_7,
\end{eqnarray}
where $f_\infty$ can go to $f_\infty=1$ with $\lambda_7,\text
{}\text{}\mu_7 \to 0$ continuously. One can turn off the coupling of
cubic term to obtain the $f_\infty$ which is defined previous
section (\ref{finfinity}).

The behavior of holographic entanglement entropy has been
studied in \cite{Hung:2011xb} for the Lovelock gravity. In this section, we
are interested in the variation of holographic entanglement entropy
after low excitation of pure AdS. For 7 dimensional Lovelock
gravity, the holographic entanglement entropy should be the minimal
value of the following functional\footnote{ There is no general
formula for HEE for higher derivative gravity as shown in
\cite{Hung:2011xb}. For the entangling surfaces with rotation
symmetry, logarithm divergent term of HEE from the (\ref{HEEF6}) is
consistent with calculation from pure CFT side. In this paper, we
will only focus on the entangling surface is 4-dimensional sphere
in (\ref{HEEF6}). Therefore, it is reasonable to use (\ref{HEEF6})
as the formula for HEE. We refer readers to \cite{Hung:2011xb} for
more general discussion.}
\begin{eqnarray}
S_A&=&\frac{2\pi}{\ell_p^5}\int_M d^5x\sqrt {h}
\Big(1+\frac{\lambda_7 L^2}{6}\mathcal{R} -\frac{\mu_7
L^4}{8}(\mathcal{R}_{\mu \nu \kappa \sigma}\mathcal{R}^{\mu \nu
\kappa \sigma}
-4\mathcal{R}_{\mu \nu}\mathcal{R}^{\mu \nu}+\mathcal{R}^2)\Big)\nonumber \\
&+&S_{\text{surfaceterm}},
 \label{HEEF6}
\end{eqnarray}
where $h$ is the determinant of the induced metric on the bulk surface $M$, $\mathcal{R}_{\mu \nu \kappa \sigma}$ is the Riemann tensor of $M$.
The surface term is added to functional to make the variational
problem work well. Following \cite{Myers:2013lva}, the surface term should be
\begin{eqnarray}
S_{\text{surfaceterm}}&=&\frac{2\pi}{\ell_p^5}\int_{\partial M} d^4x \sqrt{h_{\partial M}} \Big(\frac{\lambda_7 L^2}{3}\mathcal{K}\nonumber \\
&-&\frac{\mu_7
L^4}{8}(4\mathcal{R}^B\mathcal{K}-8\mathcal{R}^B_{ij}\mathcal{K}^{ij}-\frac{4}{3}\mathcal{K}^3+4\mathcal{K}\mathcal{K}_{ij}\mathcal{K}^{ij}-\frac{8}{3}\mathcal{K}_{ij}
\mathcal{K}^{jk}\mathcal{K}^i_k)\Big), \label{STFEE6}
\end{eqnarray}
where $\partial M$ is the boundary of $M$, $\mathcal{K}_{ij}$ and $\mathcal{K}$ are the extrinsic
curvature and the trace of extrinsic curvature on boundary $\partial
M$, $\mathcal{R}_{ij}^B$ and $\mathcal{R}^B$ are the
intrinsic Ricci tensor and Ricci scalar of the boundary $\partial M$
respectively. In the following part, we will discuss the case that
the entangling surface in the 6-dimensional CFT is a
4-dimensional sphere with radius $R_0$.
\subsection{Variance of EE after low thermal excitation}
We continue to consider low thermal excitation for 6-dimensional
CFT. For simplicity, we choose $\mu_7=-\frac{\lambda_7^2}{3}$
\footnote{Here we consider the pure AdS as a ground state of
Lovelock gravity. Generally speaking, the third order Lovelock
gravity with general parameters \cite{Dehghani:2005vh} can not admit
the black hole solution(\ref{ADSlovelock}) without the constrain
condition $\mu_7=-\frac{\lambda_7^2}{3}$. With this condition, we
can get a simple black brane solution\cite{Ge:2009ac}
\begin{eqnarray}\label{AADS7}
ds^2
=\frac{L^2}{z^2}\Big[-f(z)dt^2+\frac{dz^2}{f(z)}+\sum_{i=1}^5{dx^i}^{2}\Big],\label{ADSlovelock}
\\
f(z) =\frac{1}{\lambda_7}\bigg(1-(1-3\lambda_7\big(
1-\frac{z^6}{z_h^6}\big)^{\frac{1}{3}}\bigg), \nonumber
\end{eqnarray}
when $z_h$ is far away from boundary, we get (\ref{PIN7}).Here we
get $f_{\infty}=\frac{1-(1-3\lambda_7)^{\frac{1}{3}}}{\lambda_7}$ by
taking $\mu_7=-\frac{\lambda_7^2}{3}$ into (\ref{INFITY7}).}.

The low thermal excitation of pure AdS in 7D Lovelock gravity can
be expressed by
\begin{eqnarray}
ds^{2}=\frac{\tilde
{L}^2}{z^2}(-f(z)dt^2+\frac{dz^2}{g(z)}+dr^2+r^2d\Omega_{2}^2),
\label{PIN7}
\end{eqnarray}
with $f(z)\simeq g(z)= 1-mz^6$, $m$ is a parameter which
corresponds to the asymptotical behavior of the gravity background. The induced
metric of bulk surface $M$ is
\begin{eqnarray}
ds^2_M=\frac{\tilde
{L}^2}{z^2}\Big((1+mz^6)dz^2+dr^2+r^2d\Omega_{4}^2\Big).
\end{eqnarray}
We parameterize the surface $M$ as $r=r(u)=f(u/R_0)\cos(u/R_0)$ and
$z=z(u)=f(u/R_0)\sin(u/R_0)$, with $\epsilon \le u \le
\frac{\pi}{2}R$, and define
\begin{eqnarray}
h_{uu}=\frac{\tilde{L}^2}{z^2}\Big((1+mz^6)\dot{z}^2+\dot{r}^2\Big),
\qquad F=\log\left(\frac{r}{z}\right).
\end{eqnarray}
Taking the parametrization of the bulk surface $M$ into
($\ref{HEEF6}$), we would get a functional of $f(x)$ as follows
\begin{eqnarray}
S_A&=&\frac{2\pi\tilde{L}^5 S_4}{\ell_p^5}\int_{\epsilon
/R_0}^{\frac{\pi}{2}}dx\frac{\cos^4(x)}{\sin^5(x)}\sqrt{1+(\frac{d\log
f}{dx})^2}\Big[1+\frac{\lambda_7
 f_\infty}{6}(12\tan^2(x)+12\frac{1}{\cos^2(x)}\frac{1}{1+(\frac{d\log f}{dx})^2})\nonumber \\
&-&f_\infty^2
\mu_7\Big(3\tan^4(x)-\frac{1}{\cos^4(x)}(\frac{1}{1+(\frac{d\log
f}{dx})^2})^2+6\tan^2(x)\frac{1}{\cos^2(x)}\frac{1}{1+(\frac{d\log
f}{dx})^2}\Big)\Big].
\end{eqnarray}
The details of the calculation is given in Appendix B. With turning
off the low thermal excitation, the minimal value of functional
$S_A$ corresponds to the surface $M$:
\begin{eqnarray}
  r(u)=R_0\cos(u/R_0),\quad z(u)=R_0\sin(u/R_0),\quad \text{with} \quad \epsilon \le u \le \frac{\pi}{2}R_0.
\end{eqnarray}
As shown in Appendix A, the low thermal excitation do not affect
the configuration of bulk surface up to $O(mR_0^6)$, we can
calculate the variance of $S_A$ after the perturbation is turned on
($~\ref{PIN7}$).
\begin{eqnarray}\label{deltaS-lovelock}
\Delta S_A&=&\frac{2m R_0^6\pi\tilde{L}^5
S_4}{\ell_p^5}\int_{\epsilon /R_0}^{\frac{\pi}{2}}dx
\Big(\frac{1}{2}\sin(x)\cos^6(x)
+\lambda_7 f_\infty \sin^3(x)\cos^4(x)-\lambda_7 f_\infty \sin(x)\cos^4(x)\nonumber \\
&-&\frac{3}{2}f_\infty^2 \mu_7 \sin^5(x)\cos^2(x)-{\frac{3}{2}f_\infty^2\mu_7\sin(x)\cos^2(x)}+3f_\infty^2\mu_7\sin^3(x)\cos^2(x)\Big)\nonumber \\
&=&\frac{m R_0^6\pi\tilde{L}^5 S_4}{7\ell_p^5}(1-2\lambda_7
f_\infty-3f_\infty^2\mu_7)\nonumber\\&=&\frac{m R_0^6\pi
S_4}{7}(-{56 B_1\over 9} + 128 B_2),
\end{eqnarray}
where $S_4$ is the volume of the 4-dimensional unit sphere.
We have expressed the formula in terms of the central charges of
6-dimensional CFT in the last step (\ref{deltaS-lovelock}). Where we
have quoted the holographic expressions for the four types of
central charges denoted by $A, B_1,B_2,B_3$
\cite{Bastianelli:2000hi} \footnote{In six dimensions, the trace
anomaly can be expressed by \cite{Bastianelli:2000hi}
\begin{eqnarray} T^i_i=\sum_{i=1}^3 B_i I_i+AE_6,\nonumber
\end{eqnarray} where
\begin{eqnarray}
I_1=C_{ijkl}C^{jmnk}{{C_m}^{ il}}_{n}, \quad I_2={C_{ij}}^{  kl}{C_{kl}}^{ mn}{C_{mn}}^{  ij},\nonumber \\
I_3=C_{ijkl}(\delta_m^i\nabla^2
+4{R^i}_{m}-\frac{6}{5}R\delta^i_{m})C^{mjkl}, \quad
E_6=\frac{1}{192\pi^3}\mathcal{L}_6,\nonumber
\end{eqnarray}
where $C_{ijkl}$ is the Weyl tensor, $\mathcal {L}_6(R)$ is defined
as (\ref{EulerDensity6}).}
\begin{eqnarray}
B_1 &=& \frac{\tilde{L}^5}{\ell_p^5}\frac{-9+26f_\infty \lambda_7+51f_\infty^2\mu_7}{288},\nonumber \\
B_2 &=& \frac{\tilde{L}^5}{\ell_p^5}\frac{-9+34f_\infty \lambda_7+75f_\infty^2\mu_7}{1152},\nonumber \\
B_3 &=& \frac{\tilde{L}^5}{\ell_p^5}\frac{1-2f_\infty \lambda_7-3f_\infty^2\mu_7}{384},\nonumber \\
  A &=& \frac{\pi^3 \tilde{L}^5}{\ell_p^5}\frac{3-10f_\infty \lambda_7-45f_\infty^2\mu_7}{6}.
\end{eqnarray}

In
Appendix C we get the energy-stress tensor of CFT with the dual bulk
background (\ref{AADS7}). The variance of the energy is easy to
obtain
\begin{eqnarray} \label{deltaE-lovelock}\Delta E&=& \frac{\left(1-2f_{\infty}
\lambda_7-3\mu_7f_\infty^2\right) m R_{0}^5\tilde{L}^5 S_4 }{
2\ell_p^5}\nonumber\\&=&\frac{ m R_{0}^5 S_4 }{ 2}(-{56 B_1\over 9}
+ 128 B_2).
\end{eqnarray}
Both (\ref{deltaS-lovelock}) and (\ref{deltaE-lovelock}) show that
the higher derivative gravity will make contribution to the
variation of entanglement entropy and energy-stress tensor, as we
expected. The entanglement temperature is
\begin{eqnarray}
\frac{1}{T_{ent}}&=&\frac{2\pi}{7} R_0
\label{Tent-lovelock}
\end{eqnarray}
(\ref{deltaS-lovelock})(\ref{deltaE-lovelock}) show
that the variation of entanglement entropy and energy momentum
tensor of theory living on boundary should be related to CFT data.
The behavior of entanglement temperature (\ref{Tent-lovelock}) for
the round ball shaped subsystem calculated in both 5-dimensional and
7-dimensional Lovelock gravity cases do not depends on CFT data.
\section{Entanglement temperature in aAdS}
To extend our analysis about the entanglement temperature, we can
consider background in Fefferman-Graham coordinates corresponds to a
special excitation of the dual CFT.\\

\subsection{Asymptotical AdS in Fefferman-Graham gauge}
We assume the CFT is living on the $d$-dimensional flat space. The
entangling surface is a codimension 2 surface which is
parameterized by the coordinate $\{x_a\}$, with $a=1,2,...£¬d-2$.
And we use the Gauss coordinate, the other space coordinate is $y$, the metric is
\begin{eqnarray}
ds^2_{CFT}=-dt^2+dy^2+h_{ab}(y,x_a)dx^adx^b. \label{GC}
\end{eqnarray}
To study the asymptotical AdS space, it is convenient to use the
Fefferman-Graham coordinates in background,
\begin{eqnarray}
ds^2_{FG}=\frac{L^2}{z^2}\Big(dz^2+g_{ij}(z,x)dx^idx^j\Big).
\end{eqnarray}
It has been show that $g_{ij}(z,x)$ allow an expanding as follows
\begin{eqnarray}
g_{ij}(z,x)=g^{(0)}_{ij}(x)+g^{(2)}_{ij}(x)z^2+...+g^{(d)}_{ij}(x)z^d+\mathcal{H}z^{d}\log
z...
\end{eqnarray} Where the $d$ is even.
Via the conformal symmetry and Einstein equation the term
$g^{(2k)}_{ij}$ up to order $d-2$ and $\mathcal{H}$ can be
computable from $g^{(0)}_{ij}$ which is the dual boundary metric.
$g^{(d)}_{ij}$ is undetermined by boundary metric. Our interesting
case is the dual CFT lives on the flat spacetime, i.e.,
$g^{(0)}_{ij}=\eta_{ij}$. We consider the case that the term $g^{(2k)}_{ij}$ up to order $d-2$
and $\mathcal{H}$ are all vanishing. The asymptotical AdS is
\begin{eqnarray}
ds^2_{Bulk}=\frac{L^2}{z^2}\Big[dz^2+\Big(\eta_{ij}+\delta
\eta_{ij}(x)z^d\Big) dx^idx^j\Big].
\end{eqnarray}
Where we have denoted $\delta\eta_{ij}\equiv g^{(d)}_{ij}$. We can
use $(\ref{GC})$ and write in an explicit form as
\begin{eqnarray}
ds^2_{Bulk}=\frac{L^2}{z^2}\Big[dz^2-dt^2+\delta\eta_{ti}z^d dtdx^i+dy^2+\delta\eta_{yi}z^ddydx^i+(h_{ab}+\delta\eta_{ab}z^d)dx^a dx^b\Big].\nonumber \\
\label{AASDGC}
\end{eqnarray}
We consider general low excitation of the pure AdS gravity would
correspond to (\ref{AASDGC}). We can use the holographic method to
analyze the variance of EE and relation with variance of the energy
of the subsystem.

\subsection{Variance of EE and energy}
When the excitation is turned off, i.e., the background should go back
to pure AdS. We can get the entanglement entropy
of the subsystem A by minimizing the functional of bulk
surface $M$, which is parameterized as $z=z(y)$. Here we have
assumed the entangling surface has the symmetry to make the
parametrization possible. The induced metric $g_{\mu\nu}$ on $M$
is
\begin{eqnarray}
ds^2_M=\frac{L^2}{z^2}\Big((1+z'^2)dy^2+h_{ab}dx^adx^b\Big),
\label{PADSM}
\end{eqnarray}
where $z'\equiv \frac{dz}{dy}$.
\begin{eqnarray}\label{PEE}
S=\frac{2\pi}{\ell_p^{d-1}}\int_M
dyd^{d-2}x\sqrt{g}=\frac{2\pi L^{d-1}}{\ell_p^{d-1}}\int_M
dyd^{d-2}x\frac{\sqrt{1+z'^2}}{z^{d-1}}\sqrt{h},
\end{eqnarray}
with $h\equiv \det\ h_{ab}$. We can find the minimal surface $M_{0}$
with a solution $z=z_{0}(y)$. When the excitation is
turned on, the entanglement entropy would also make a change and
the induced metric on $M$ would become
$\tilde{g}_{\mu\nu}=g_{\mu\nu}+\delta g_{\mu\nu}$,
\begin{eqnarray}
d\tilde{s}^2_M=\frac{L^2}{z^2}\Big((1+z'^2+\delta
\eta_{yy}z^d)dy^2+(h_{ab}+\delta \eta_{ab}z^d)dx^a
dx^b+\delta\eta_{ya}z^d dy dx^a\Big), \label{ASM}
\end{eqnarray}
We use statement in Appendix A to get the variance of holographic
entanglement entropy. The perturbation does not change the shape of
$M$ up to $O(\delta g_{\mu\nu})$. So we just need expand the
functional
\begin{eqnarray}
\tilde{S}=\frac{2\pi}{\ell_p^{d-1}}\int_M dyd^{d-2}x \sqrt{\tilde{g}}.
\end{eqnarray}
to $O(\delta g_{\mu\nu})$ and keep the minimal surface to be $M_0$
with $z=z_{0}(y)$.
\begin{eqnarray}
\tilde{S}=\frac{2\pi}{\ell_p^{d-1}}\int_{M_0}dyd^{d-2}x\Big(\sqrt{g}(1+\frac{1}{2}g^{\mu\nu}\delta
g_{\mu\nu}+O(\delta g_{\mu\nu}^2))\Big)
\end{eqnarray}
Then we get
\begin{eqnarray}
\Delta S=
\tilde{S}-S=\frac{2\pi}{\ell_p^{d-1}}\int_{M_0}dyd^{d-2}x\Big(\frac{1}{2}\sqrt{g}g^{\mu\nu}\delta
g_{\mu\nu}+O(\delta g_{\mu\nu}^2))\Big).
\end{eqnarray}
We can read $\delta g_{\mu\nu}$ by comparing ($\ref{ASM}$) with
($\ref{PADSM}$). The result is
\begin{eqnarray}
\Delta S=\frac{\pi
L^{d-1}}{\ell_p^{d-1}}\int_{M_0}dyd^{d-2}x\sqrt{h}\sqrt{1+z_{0}'^2}z_0\Big(\frac{\delta
\eta_{yy}}{1+z_{0}'^2}+h^{ab}\delta\eta_{ab}\Big). \label{VOEE}
\end{eqnarray}\\
Now we turn to calculate the variance of energy of the subsystem after the low
excitation of the ground state (pure AdS). The energy-stress tensor
for the CFT living on the boundary is
\begin{eqnarray}
T_{ij}=\frac{d L^{d-1}}{\ell_p^{d-1}}\delta\eta_{ij} \label{EST}
\end{eqnarray}
The conformal symmetry would restrict the energy-stress tensor to be
traceless, i.e., $T_{ij}\eta^{ij}=0$. According to $(\ref{EST})$
$\delta\eta_{ij}$ would also restrict to be traceless, in the
coordinate $(\ref{GC})$ we get
\begin{eqnarray}
-\delta \eta_{tt}+\delta \eta_{yy}+h^{ab} \delta \eta_{ab}=0.
\label{ETTL}
\end{eqnarray}
The variance of energy of the subsystem A is
\begin{eqnarray}
\Delta E= \int_A dyd^{d-2}x  \sqrt{h} T_{tt}=\frac{d
L^{d-1}}{\ell_p^{d-1}}\int_A dyd^{d-2}x \sqrt{h} \delta\eta_{tt}.
\label{GCVOE}
\end{eqnarray}
Using $(\ref{EST})$$(\ref{ETTL})$, we rewritten $(\ref{VOEE})$ as
\begin{eqnarray}
\Delta S=\frac{\pi}{d}\int_{A}dyd^{d-2}x\sqrt{h}\sqrt{1+z_{0}'^2}z_0\Big(T_{tt}-\frac{z_{0}'^2}{1+z_{0}'^2}T_{yy}\Big). \label{GCFR}
\end{eqnarray}
The variance of holographic entanglement entropy ($\ref{GCFR}$)  of
the subsystem $A$ is only related to $T_{tt}$ and $T_{yy}$, i.e.,
the variance energy density and the strength of pressure in the
direction normal to the entangling surface.

\subsection{Entanglement temperature with sphere profile}
In this subsetion, we would like to consider the subsystem A to be
an $(d-2)$-dimensional sphere with radius $R_0$ as a example to
investigate the entanglement temperature. We get
$h_{ab}dx^adx^b=y^2d\Omega^2_{d-2}$. It is easy to get the minimal
surface $M_{0}$, $z_{0}=\sqrt{R_0^2-y^2}$, with $0\le y\le R_0$.
Taking this solution into $(\ref{GCFR})$, we get
\begin{eqnarray}
\Delta S=\frac{\pi }{d}\int_{A}dyd^{d-2}x \sqrt{h}
\Big(R_0 T_{tt}-\frac{y^2}{R_0}T_{yy}\Big)
\end{eqnarray}
We can rewrite the above formula in a more compact form by using
$(~\ref{GCVOE})$,
\begin{eqnarray}
\Delta S=\frac{\pi R_0}{d}\Delta E-\pi\int_A
dyd^{d-2}x\sqrt{h}\frac{y^2}{R_0}T_{yy}. \label{NR}
\end{eqnarray}

We find that the second term on the right hand side of ($\ref{NR}$)
is not linear dependence on $\Delta E$ generally. There is
possibility to break first law-like relation in some special cases.
We do not consider the dynamic constrain \cite{Nozaki:2013vta} of
the perturbation of the bulk background in this paper. At this
stage, we have no exact statement about whether the first law-like
relation will be hold or not. The (\ref{NR}) shows that $\Delta S$
highly depend on the configurations of subsystem and dynamic
constrain of the perturbation.

In the low thermal excitation, $\delta\eta_{tt}$ and
$\delta\eta_{yy}$ are constant. Here we take them to be function of the coordinate $y$ which is normal
to the sphere. Generally speaking, we assume
\begin{eqnarray}
\delta\eta_{tt}=\sum_{i}m_{(i)}(\frac{y}{R_0})^{i}, \label{GYT}\\
\delta\eta_{yy}=\sum_{i}n_{(i)}(\frac{y}{R_0})^{i}. \label{GYY}
\end{eqnarray}
The parameters $m_{(i)},n_{(i)} \ll 1$. Taking $(\ref{GYT})$ into
($\ref{GCVOE}$), we get
\begin{eqnarray}
\Delta E=\frac{dL^{d-1}R_0^{d-1}S_{d-2}}{2\ell_p^{d-1}}
\sum_{i}\frac{m_{(i)}}{d+i-1}. \label{EFSVOE}
\end{eqnarray}
Taking $(\ref{GYT})$ and $(\ref{GYY})$ into $(\ref{VOEE})$, we get
$\Delta S$,
\begin{eqnarray}
\Delta S=\frac{2\pi R_0}{d} \Delta E - \frac{\pi
L^{d-1}R_0^{d}S_{d-2}}{\ell_p^{d-1}}\sum_{i}\frac{n_{(i)}}{d+i+1}.
\end{eqnarray}
Or by using ($\ref{EFSVOE}$),
\begin{eqnarray}
\Delta S=\frac{2\pi R_0}{d}\Big(1-\frac{A}{B}\Big)\Delta E,
\label{RFE}
\end{eqnarray}
with
\begin{eqnarray}
A=\sum_{i}\frac{n_{(i)}}{d+i+1},\quad B=
\sum_{i}\frac{m_{(i)}}{d+i-1}.
\end{eqnarray}
In this example, the $\Delta S$ is also proportional to the variance
$\Delta E$ and the entanglement temperature is related to $m_{(i)}$
and $n_{(i)}$, which correspond to the low excitation mode. When
$n_{{i}}=m_{(i)}=0$, for $i\ge 1$, and $n_{(0)}=\frac{m}{d},
m_{(0)}=\frac{(d-1)m}{d}$, ($\ref{RFE}$) recover the result of
\cite{Bhattacharya:2012mi}, i.e., $\Delta S=\frac{2\pi
R_0}{d+1}\Delta E$.

This example shows that the first law-like relation and
entanglement temperature is not so obvious that it will need to be
more carefully study. It is worth to study entanglement temperature
in subsystem with general configurations and excitations.
\section{Conclusion and Discussion}
The entanglement entropy is a useful quantity to describe the
entanglement structure of the CFT in vacuum state. It is also
interesting to study the property of the entanglement entropy after
low excitation of the vacuum state. In this paper, we mainly focus
on the entanglement entropy property of the low excited state in higher derivative gravity by
using the holographic method. The variance
of the HEE would contain more information of the CFT after the
low thermal excitation. The entanglement temperature will
absorb the contribution from higher derivative gravity.  When assuming the bulk theories are Lovelock
gravity, one considers these solutions are the low excited state of
vacuum of CFT. We just turn on the low thermal excitation and we get
the variance of the entanglement entropy for sphere and infinite
stripe profile. Combing the variation of energy of these subsystem
on the boundary, one can find a novel relation which is first law-like theorem for entanglement entropy. Finally, we also obtain the
entanglement temperature which is similar to
\cite{Bhattacharya:2012mi} and the coefficient absorbs the
contribution of higher derivative gravity. We find
that the entanglement temperature should depend on the
central charges of corresponding CFT in strip shaped subsystem in
5-dimensional Lovelock gravity. The behavior of entanglement
temperature for the round ball shaped subsystem calculated in both
5-dimensional and 7-dimensional Lovelock gravity cases do not
depends on CFT data.

Finally, we study more general low excited states in the
Fefferman-Graham coordinates. We get the general result of variance
of entanglement entropy $(\ref{GCFR})$, which is related with the
$t-t$ and $y-y$ component of energy-stress tensor. The variance of
the entanglement entropy with the energy density and pressure which
is normal to the entangling surface in general excited states. At
this stage, we can not conclude whether the first law-like relation
will be hold or not. We have show (\ref{NR}) shows that $\Delta S$
highly depend on the configurations of subsystem and dynamic
constrain of the perturbation. We also take a subsystem A with a
$(d-2)$-dimensional sphere as an example to show the entanglement
temperature and the first law-like relation. This may give us more
clues to study the quantum entanglement structure of the low excited
CFT and more insight to the probable first law relation of
entanglement entropy.

In the future, we would like to study the entanglement temperature
and entanglement density in general higher derivative gravity.
Although there is no universal formula to count the entanglement
entropy in general higher derivative gravities, it is worth to try
and check whether the first and second law like theorem are correct.
One more project is to study the HEE and entanglement temperature of
subsystem with other general configurations. Finally, authors
\cite{Li:2011hp}\cite{He:2011hw}\cite{Cai:2012xh}\cite{Cai:2012eh}
have used potential construction approach to generate some gravity
solutions analytically. These solutions are good place to study the
dynamics of entanglement entropy \cite{Nozaki:2013vta}.

\section*{Acknowledgements}
The authors are grateful to Ronggen Cai, Danning Li, Li Li, Miao Li,
Tianjun Li, Junbao Wu and Haitang Yang for useful conversations and
correspondence. WG especially thanks Miao Li for supports and
encouragements. Further we should thank for Tadashi Takayanagi's
nice suggestions and comments on this version. SH are
supported in part by the National Natural Science Foundation of
China (No.10821504, No.10975168 and No.11035008, No. 11147106), and
in part by Shanghai Key Laboratory of Particle Physics and Cosmology
under grant No.11DZ2230700. WG  are supported by the National
Natural Science Foundation of China (Grant Nos.~11275247 and
10821504). WG and SH would like to appreciate ``The
7th Asian Winter School on Strings, Particles and Cosmology " hold
in Beijing, China for financial support. SH would like to thank the
``Spring School on Superstring Theory and Related Topics" held at
ICTP, Italy for their hospitality and financial support. SH also
would like appreciate the general financial support from China
Postdoctoral Science Foundation No. 2012M510562.

\appendix
\renewcommand{\theequation}{\thesection.\arabic{equation}}
\addcontentsline{toc}{section}{Appendices}
\section*{APPENDIX}
\section{Variation of HEE }
In this appendix, we will give a brief proof that the method to
calculate the variance of HEE with turning on low excitation is
reasonable. The basic logic is that the low excitation of the CFT
corresponds to turning on small perturbation term related to $m_i$
in the pure AdS background. The perturbation is described by some dimensionless
parameters $m_1...m_i...m_k$, which are quite small, i.e., $m_i\ll
1$$(1\le i\le k)$. We only care about the leading effect of $m_i$ in
changing the classical configuration of entangling surface. The
holographic entanglement entropy is given by minimizing the
functional $S_A$ of entanglement surface $M$, which is parameterized
by $z=z(x,...,y)$, where $x,...,y$ are the boundary coordinates.
When the perturbation is turn off, the background goes back to pure
AdS, we assume that the minimal $S_A(z_0, m_i=0)$ corresponds to the
surface $m$ $z=z_0(x,...,y)$. Then
\begin{eqnarray}
\frac{\delta S_A}{\delta z}\arrowvert_{z=z_0}=0 \label{Variance}
\end{eqnarray}
with turning on the perturbation, $S_A$ is still functional of the
bulk surface with the parameters $m_i$$(1\le i\le k)$, i.e.,
$\tilde{S}_A\equiv S_A(z;m_1,...,m_k)$. In principle, we can
variance $S_A$ and get the minimal value with a solution
$\tilde{z}\equiv z(x,...,y; m_1,...,m_k)$. With assuming that the
perturbation is quite small, we can expand $z=z(x,...,y;
m_1,...,m_k)$ up to first order of the parameters $ m_i$$(1\le i\le
k)$.
\begin{eqnarray}
z(x,...,y; m_1,...,m_k)=z_0(x,...,y)+\delta z(x,...,y;m_1,...,m_k).
\end{eqnarray}
We have denoted $\delta z(x,...,y;m_1,...,m_k)=m_iz^{(i)}(x,...,y)$.
We can also expand $\tilde{S}_A$ as
\begin{eqnarray}
\tilde{S}_A\equiv S_A(z;m_1,...,m_k)=S_A(z, m_i=0)+m_iS^{(i)}(z).
\end{eqnarray}
After low thermal excitation, the holographic entanglement entropy
should be
\begin{eqnarray}
S_A(\tilde{z};m_1,...,m_k)&=&S_A(z_0(x,...,y)+\delta z(x,...,y;m_1,...,m_k);m_1,...,m_k)\nonumber \\
&=&S_A(z_0(x,...,y)+\delta z(x,...,y;m_1,...,m_k))\nonumber \\
 &+&m_iS^{(i)}(z_0(x,...,y)+\delta z(x,...,y;m_1,...,m_k))+O(m_i^2)\nonumber \\
&=&S_A(z_0(x,...,y))+\frac{\delta S_A}{\delta z}\arrowvert_{z=z_0}\delta z(x,...,y;m_1,...,m_k)+m_iS^{(i)}(z_0)+O(m_i^2)\nonumber \\
&=&S_A(z_0(x,...,y))+m_iS^{(i)}(z_0)+O(m_i^2).
\end{eqnarray}
Where we have used ($~\ref{Variance}$). Finally, the variance of the
holographic entanglement entropy $\Delta
S_A=m_iS^{(i)}(z_0)+O(m_i^2)$. In other word, the variance of the
shape of the bulk surface do not give contribution to
variance of the functional $S_A$ up to $O(m_i)$.

\section{Surface term}
In this subsection, we will list some details about the surface term
which make the variation to be well defined. The bulk surface $M$ is
parameterized as $r=r(u)$ and $z=z(u)$, with $u_i \le u \le u_f$. So
the boundary $\partial M$ is the hypersurface $u=u_i$. Using the
induced metric on $M$, we can get
\begin{eqnarray}
&&\mathcal{R}_{\mu \nu \kappa \sigma}\mathcal{R}^{\mu \nu \kappa \sigma}=\frac{1}{6}(\mathcal{R}^B-12h^{uu}\dot{F}^2)^2+16(\Delta_u F+h^{uu}\dot{F}^2)^2,\nonumber \\
&&\mathcal{R}_{\mu \nu}\mathcal{R}^{\mu \nu}=\frac{1}{4}\Big(\mathcal{R}^B-4(\Delta_u F+4h^{uu}\dot{F}^2)\Big)^2+16(\Delta_u F+h^{uu}\dot{F}^2)^2,\nonumber \\
&&\mathcal{R}^2=(\mathcal{R}^B-8\Delta_u F-20h^{uu}\dot{F}^2)^2.
\label{STOEE}
\end{eqnarray}
Where $\Delta_u F \equiv \frac{1}{\sqrt{h_{uu}}}\partial_u
\frac{\dot{F}}{\sqrt{h_{uu}}}$. The boundary is a 4-dimensional
sphere with radius $e^{2F}\tilde{L}^2$, we get
$\mathcal{R}^B_{ij}=\frac{3h_{ij}}{e^{2F}\tilde{L}^2}$ and
$\mathcal{R}^B=\frac{12}{e^{2F}\tilde{L}^2}$, where $h_{ij}$ is the
induced metric on the boundary $\partial M$. The normal outward unit
vector to $\partial M$ is $n_\nu=-\sqrt{h_{uu}}\delta_{u\nu}$. It is
easy to get the extrinsic curvature $\mathcal{K}_{\nu
\mu}=\bigtriangledown_\nu n_\mu\arrowvert_{u=u_i}$ and its trace,
where $\bigtriangledown $ is defined on the bulk surface $M$
can be obtain as follows
\begin{eqnarray}
\mathcal{K}_{uu}=\mathcal{K}_{ui}=\mathcal{K}_{iu}=0,\quad
\mathcal{K}_{ij}=
-\frac{\dot{F}h_{ij}}{\sqrt{h_{uu}}}\arrowvert_{u=u_i}.
\end{eqnarray}
The surface term ($~\ref{STFEE6}$) are listed
\begin{eqnarray}
&&\mathcal{K}=-\frac{4\dot{F}}{\sqrt{h_{uu}}}\arrowvert_{u=u_i},\quad \mathcal{R}^B\mathcal{K}=-e^{-2F}\frac{48\dot{F}}{\tilde{L}^2\sqrt{h_{uu}}}\arrowvert_{u=u_i},\quad \mathcal{R}^B_{ij}\mathcal{K}^{ij}=-e^{-2F}\frac{12\dot{F}}{\tilde{L}^2\sqrt{h_{uu}}}\arrowvert_{u=u_i},\nonumber \\
&&\mathcal{K}\mathcal{K}_{ij}\mathcal{K}^{ij}=-16\Big(\frac{\dot{F}}{\sqrt{h_{uu}}}\Big)^3\arrowvert_{u=u_i},\quad
\mathcal{K}_{ij}\mathcal{K}^{jk}\mathcal{K}^i_k)=-4\Big(\frac{\dot{F}}{\sqrt{h_{uu}}}\Big)^3\arrowvert_{u=u_i}.
\end{eqnarray}
The surface term can be expressed
\begin{eqnarray}
S_{\text{surfaceterm}}=\frac{2\pi}{\ell_p^5}\int d\Omega_4\tilde{L}^4e^{4F} \Big(&-&\frac{4\lambda_7 L^2}{3}\frac{\dot{F}}{\sqrt{h_{uu}}}\nonumber \\
&-&\frac{\mu_7
L^4}{8}(-e^{-2F}\frac{96\dot{F}}{\tilde{L}^2\sqrt{h_{uu}}}+32(\frac{\dot{F}}{\sqrt{h_{uu}}})^3\Big)\arrowvert_{u=u_i}.
\nonumber \\
=\frac{2\pi S_4 \tilde{L}^4}{\ell_p^5}e^{4F}\Big(&-&\frac{4\lambda_7 L^2}{3}\frac{\dot{F}}{\sqrt{h_{uu}}}\nonumber \\
&-&\frac{\mu_7
L^4}{8}(-e^{-2F}\frac{96\dot{F}}{\tilde{L}^2\sqrt{h_{uu}}}+32(\frac{\dot{F}}{\sqrt{h_{uu}}})^3\Big)\arrowvert_{u=u_i}.
\end{eqnarray}
Where $S_4$ is the volume of the 4-dimensional unit sphere. Using
($~\ref{STOEE}$), we get ($~\ref{HEEF6}$) as
\begin{eqnarray}
S_{A}&=&\frac{2\pi\tilde{L}^4
S_4}{\ell_p^5}\int_{u_i}^{u_f}du\sqrt{h_{uu}}e^{4F}
\Big[1+\frac{\lambda L^2}{6}((\mathcal{R}^B-8\Delta_u F-20h^{uu}\dot{F}^2))\nonumber \\
&-&\frac{\mu L^4}{8}(e^{4F}\frac{24}{\tilde{L}^4}+120(h^{uu}\dot{F}^2)^2-e^{-2F}\frac{144h^{uu}\dot{F}^2}{\tilde{L}^2}+96\Delta_u F h^{uu}\dot{F}^2-e^{-2F}\Delta_u F \frac{96}{\tilde{L}^2})\Big]\nonumber \\
&&\qquad +S_{\text{surfaceterm}}.
\end{eqnarray}
By integrating by parts, we get
\begin{eqnarray}
S_A&=&\frac{2\pi\tilde{L}^4 S_4}{\ell_p^5}\int_{u_i}^{u_f}du\sqrt{h_{uu}}e^{4F}\Big[1+\frac{\lambda_7 L^2}{6}(e^{-2F}\frac{12}{\tilde{L}^2}+12h^{uu}\dot{F}^2)\nonumber \\
&-&L^4\mu_7\Big(e^{-4F}\frac{3}{\tilde{L}^4}
-(h^{uu}\dot{F}^2)^2+e^{-2F}\frac{6}{\tilde{L}^2}h^{uu}\dot{F}^2\Big)\Big]-\frac{8\pi\lambda_7 L^2\tilde{L}^4 S_4}{3\ell_p^5}e^{4F}\frac{\dot{F}}{\sqrt{h_{uu}}}\arrowvert_{u=u_i}^{u=u_f}\nonumber \\
&-&\frac{2\pi\mu_7L^4\tilde{L}^4
S_4}{\ell_p^5}\Big(4e^{4F}(\frac{\dot{F}}{\sqrt{h_{uu}}})^3
-e^{2F}\frac{12}{\tilde{L}^2}\frac{\dot{F}}{\sqrt{h_{uu}}}\Big)\arrowvert_{u=u_i}^{u=u_f}+S_{\text{surfaceterm}}.
\end{eqnarray}
As we can see the surface term exactly cancel the boundary
contribution from the integration by parts. $S_A$ is
\begin{eqnarray}
 S_A&=&\frac{2\pi\tilde{L}^4 S_4}{\ell_p^5}\int_{u_i}^{u_f}du\sqrt{h_{uu}}e^{4F}\Big[1+\frac{\lambda_7 L^2}{6}(e^{-2F}\frac{12}{\tilde{L}^2}+12h^{uu}\dot{F}^2)\nonumber \\
&-&L^4\mu_7\Big(e^{-4F}\frac{3}{\tilde{L}^4}
-(h^{uu}\dot{F}^2)^2+e^{-2F}\frac{6}{\tilde{L}^2}h^{uu}\dot{F}^2\Big)\Big].
\end{eqnarray}
To minimal functional, one can obtain the following configuration
which can be parameterized $r$ and $z$ as
\begin{eqnarray}
r(u)=f(u/R_0)\cos(u/R_0),\quad z(u)=f(u/R_0)\sin(u/R_0),\quad with
\quad \epsilon \le u \le \frac{\pi}{2}R_0.
\end{eqnarray}
One can find that $S_A$ is a functional of $f(x)$:
\begin{eqnarray}
S_A&=&\frac{2\pi\tilde{L}^5 S_4}{\ell_p^5}\int_{\epsilon /R_0}^{\frac{\pi}{2}}dx
\frac{\cos^4(x)}{\sin^5(x)}\sqrt{1+(\frac{d\log f}{dx})^2}\Big[1+\frac{\lambda_7 f_\infty}{6}(12\tan^2(x)+12\frac{1}{\cos^2(x)}\frac{1}{1+(\frac{d\log f}{dx})^2})\nonumber \\
&-&f_\infty^2
\mu_7\Big(3\tan^4(x)-\frac{1}{\cos^4(x)}(\frac{1}{1+(\frac{d\log
f}{dx})^2})^2+6\tan^2(x)\frac{1}{\cos^2(x)}\frac{1}{1+(\frac{d\log
f}{dx})^2}\Big)\Big].
\end{eqnarray}
From the above action, the equation of motion can be obtain \bea
0&=&\cot ^4(x) \csc (x) G''(x) \left(2 \lambda  f_{\infty }+3 \mu _7
f_{\infty }^2-1\right)+\cot (x) \csc (x) G'(x)^7 \Big(f_{\infty }
\left(4 \lambda  \csc ^2(x)-3 \mu _7 f_{\infty
}\right)\nonumber\\&+&2 \cot ^2(x) \left(\lambda  f_{\infty }+2 \csc
^2(x)\right)+\cot ^4(x)\Big)+\frac{1}{2} (\cos (2 x)+9) \cot ^3(x)
\csc ^3(x) G'(x) \Big(-2 \lambda  f_{\infty }\nonumber\\&-&3 \mu _7
f_{\infty }^2+1\Big)+\frac{1}{8} \cot (x) \csc ^5(x) G'(x)^5 \Big(-2
\lambda  f_{\infty }+45 \mu _7 f_{\infty }^2-12 \cos (2 x) \left(6
\lambda  f_{\infty }+3 \mu _7 f_{\infty
}^2-5\right)\nonumber\\&-&\cos (4 x) \left(6 \lambda  f_{\infty }+9
\mu _7 f_{\infty }^2-3\right)+57\Big)\nonumber\\&+&\frac{1}{8} \cot
(x) \csc ^5(x) G'(x)^3 \Big(-58 \lambda  f_{\infty }-3 \mu _7
f_{\infty }^2-\cos (4 x) \left(6 \lambda  f_{\infty }+9 \mu _7
f_{\infty }^2-3\right)\nonumber\\&-&12 \cos (2 x) \left(8 \lambda
f_{\infty }+9 \mu _7 f_{\infty
}^2-5\right)+57\Big)\nonumber\\&+&\cot ^2(x) \csc ^3(x) G'(x)^2
G''(x) \left(-4 \lambda  f_{\infty }-15 \mu _7 f_{\infty }^2+\cos (2
x) \left(2 \lambda  f_{\infty }+3 \mu _7 f_{\infty
}^2-1\right)-1\right)\nonumber\\&+&\cot (x) G'(x)^4 G''(x)
\Big(\lambda f_{\infty } (\cos (2 x)-5) \cot (x) \csc
^3(x)\nonumber\\&+&3 \mu _7 f_{\infty }^2 \left(4 \csc
^2(x)+1\right) \sec (x)+\cot ^3(x) (-\csc (x))\Big)\eea with
$G(x)=\log f(x)$. Where the prime is derivative with respect to $x$.
The Euler-Lagrange equation allow a solution that
$f=\text{Constant}$. The minimal value of functional $S_A$
corresponds to the surface $M$ classically.
\begin{eqnarray}
  r(u)=R_0\cos(u/R_0),\quad z(u)=R_0\sin(u/R_0),\quad \text{with} \quad \epsilon \le u \le \frac{\pi}{2}R_0.
\end{eqnarray}
Where we have used boundary condition $f(u/R_0)=R_0$.

\section{Energy momentum tensor on the boundary theory}
In this section, we would like to deal with the energy momentum
tensor of CFT when the dual gravity are 5-dimensional AdS
GB gravity and 7-dimensional Lovelock gravity respectively. To
define the proper boundary stress-energy tensor, the total
gravitational action should have following contribution from three
parts\cite{KS}\cite{SKS}
\begin{eqnarray}
I=I_{\text{bulk}}(g_{ij})+I_{\text{surf}}(g_{ij})+I_{\text{ct}}(\gamma_{ij}).
\end{eqnarray}
$I_{\text{bulk}}(g_{ij})$ denotes for the bulk dynamics,
$I_{\text{surf}}(g_{ij})$ denotes for surface terms contribution and
$I_{\text{ct}}(\gamma_{ij})$ denotes the terms to make the total
action to be finite.

\subsection{5-dimensional Lovelock gravity}
The 5-dimensional gravity action with the Gauss-Bonnet term in the
bulk $M$ with a boundary $\partial M$ is\cite{Cvetic:2001bk}\cite{Lidsey:2002zw}\cite{Dehghani:2006dh}
\begin{eqnarray}\label{C2}
I_{ren}&=&\frac{1}{2\ell_p^3} \int_M d^5x  \sqrt{-g} [
\frac{12}{L^2} + R + \frac{\lambda_5 L^2}{2} L_4 ]\nonumber \\&-&\frac{1}{2{\ell_p}^3} \int_{\partial M}d^4x\sqrt{-\gamma}\Big[2K-{3\over L}+2\lambda_5 L^2\Big(J-(R_{ij}-\frac{1}{2}R\gamma_{ij})K^{ij}\Big)\Big]+I_{ct},\nonumber \\
 \end{eqnarray}
 with
 \begin{eqnarray}
 J_{ij}=\frac{1}{3}(2KK_{ik}K^{k}_j+K_{kl}K^{kl}K_{ij}-2K_{ik}K^{kl}K_{lj}-K^2K_{ij}),\quad J=J_{ij}\gamma^{ij}.
 \end{eqnarray}
Where $K_{ij}$ and $K$ are respective the extrinsic curvature and
its trace of the boundary $\partial M$, $\gamma_{ij}$ is the induced
metric on the boundary $\partial M$, $R_{ij}$ and $R$ are respective
the induced Ricci tensor and Ricci scalar of the boundary $\partial
M$. The metric of the bulk theory is ($\ref{AADS}$). To regulate the
theory, we restrict to the region $z\ge \epsilon$ and the surface
term is evaluated at $z=\epsilon$. As the boundary is flat the term
$R_{ij}$ and $R$ do not contribute. The induced metric
$\gamma_{ij}=\frac{\tilde{L}^2}{\epsilon^2} g_{ij}(x,\epsilon)$,
where the leading term of $g_{ij}(x,\epsilon)$ expanded as
$\epsilon$ is the flat metric $g_{(0)}^{ij}$. Then the one point
function of stress-energy tensor of the dual CFT is given by
\cite{KS}\cite{SKS}
\begin{eqnarray}
T_{ij}=\frac{2}{\sqrt{-\det g_{(0)}}}\frac{\delta I_{ren}}{\delta
g_{(0)}^{ij}}=\lim_{\epsilon \to 0}\Big(\frac{\tilde{L}^2}{\epsilon^2}\frac{2}{\sqrt{-\gamma}}\frac{\delta I_{ren}}{\delta
\gamma^{ij}}\Big). \label{CFTET}
\end{eqnarray}
Two terms will contribute to the finite part of boundary energy-stress tensor according to (\ref{C2}),
one is from the $O(\epsilon^2)$ of the Brown-York tensor $T^{BY}_{ij}$ on the boundary $z=\epsilon$, with
\begin{eqnarray}\label{BY}
T^{BY}_{ij}=-\frac{1}{\ell_p^3}\Big[(K_{ij}-K\gamma_{ij})+\lambda_5
L^2(3J_{ij}-J\gamma_{ij})\Big],
\end{eqnarray}
other one is from the counter term. In our case, the counter term is very simply for the boundary is flat. There is only one necessary counter term
\begin{eqnarray}\label{CT}
I_{ct}=-\frac{1}{\ell_p^3}\int_{\partial
M}d^dx\frac{(d-1)\sqrt{-\gamma}}{L'},
\end{eqnarray}
where $L'$ depends on $L$ and Gauss-Bonnet parameter $\lambda_5$.
The explicit formula about $L'$ for (\ref{C2}) is \bea L' ={3
\tilde{L}^3 L\over-3 \tilde{L}^3 + 3 \tilde{L}^2L + 2 L^3
\lambda_5}. \eea Directly evaluate (\ref{BY}) using (\ref{AADS}), we
get \bea T_{tt}=\frac{3 m \tilde{L}^3(1-2 \lambda_5 f_\infty)}{2
\ell_p^3} . \label{ETGB}\eea We can see that the contribution tensor
from Gauss bonet gravity is related to the term which is
proportional to $\lambda$. If one turns off the Gauss-Bonnet
correction, the energy density can be reduced to that of
CFT with the dual pure Einstein gravity\cite{Bhattacharya:2012mi}.

\subsection{7-dimensional Lovelock gravity}
In this subsection, we will deal with energy momentum tensor in the
6-dimensional CFT with the dual 7-dimensional Lovelock gravity. We
recall that the gravity action should be\cite{Dehghani:2006dh}
\begin{eqnarray}\label{C9}
 I_{ren}&=&\frac{1}{2\ell_p^5}\int d^{7}x \sqrt{-g} (\frac{30}{L^2}+R+\frac{L^2}{12}\lambda_7 \mathcal{L}_4(R)-\frac{L^4}{24}\mu_7 \mathcal{L}_6(R))\nonumber \\
 &-&\frac{1}{2\ell_p^5}\int d^6x\sqrt{-\gamma}\Big[2K-{5\over L}+\frac{\lambda_7 L^2}{3}\Big(J-2(R_{ij}-\frac{1}{2}R\gamma_{ij})K^{ij}\Big)\nonumber \\
 &-&\frac{\mu_7 L^4}{4}\Big(P-2\tilde{G}_{ij}K^{ij}-12R_{ij}J^{ij}+2RJ-4KR_{ijkl}K^{ik}K^{jl}-8R_{ijkl}K^{ik}K^j_mK^{ml}\Big)\Big]\nonumber
 \\ &+&I_{ct}.
\end{eqnarray}
with
\begin{eqnarray}
P_{ij}&=&\frac{1}{5}\Big[\Big(K^4-6K^2K_{mn}K^{mn}+8KK_{mn}K^n_kK^{km}-6K_{mn}K^{nk}K_{kl}K^{lm}+3(K_{mn}K^{mn})^2\Big)K_{ij}\nonumber \\
&-&(4K^3-12KK_{kl}K^{kl}+8K_{mn}K^n_lK^{lm})K_{ik}K^{k}_j-24KK_{il}K^{lk}K_{km}K^m_j \nonumber \\
&+&12(K^2-K_{mn}K^{mn})K_{il}K^{lk}K_{kj}+24K_{ik}K^{kl}K_{lm}K^{mn}K_{nj}\Big],\quad
P=\gamma_{ij}P^{ij}
\end{eqnarray}
and
\begin{eqnarray}
\tilde{G}_{ij}=2(R_{ikmn}R_j^{\ kmn}-2R_{ikjl}R^{kl}-2R_{ik}R^k_j+RR_{ij})\nonumber \\
-\frac{1}{2}(R_{ijkl}R^{ijkl}-4R_{ij}R^{ij}+R^2)\gamma_{ij}.
\end{eqnarray}
we can get the dual CFT stress-energy tensor using the same
procedure mentioned in above section, (\ref{CFTET}) should be
\begin{eqnarray}
T_{ij}=\frac{2}{\sqrt{-\det g_{(0)}}}\frac{\delta I_{ren}}{\delta
g_{(0)}^{ij}}=\lim_{\epsilon \to 0}\Big(\frac{\tilde{L}^4}{\epsilon^4}\frac{2}{\sqrt{-\gamma}}\frac{\delta I_{ren}}{\delta
\gamma^{ij}}\Big). \label{CFTET7}
\end{eqnarray}
The Brown-York tensor is
\begin{eqnarray}
T^{BY}_{ij}=-\frac{1}{\ell_p^5}\Big[(K_{ij}-K\gamma_{ij})+\frac{\lambda_7
L^2}{6}(3J_{ij}-J\gamma_{ij})-\frac{\mu_7
L^4}{8}(5P_{ij}-P\gamma_{ij})\Big].
\end{eqnarray}
The counter term is still as the form (\ref{CT}), but $L'$ should
also depend on the new parameters $\lambda_7$ and $\mu_7$. The
explicit formula about $L'$ for (\ref{C9}) is \bea L' ={ 15
\tilde{L}^5 L\over-15 \tilde{L}^5 + 15 \tilde{L}^4L + 10 \tilde{L}^2
L^3 \lambda_7 + 9 L^5 \mu_7}. \eea In our special asymptotically
$AdS_7$ background ($~\ref{PIN7}$), the boundary is regularized at
$z=\epsilon$, the boundary is flat, we obtain $t-t$ component of the
energy-stress tensor,
\begin{eqnarray}
 T_{tt}&=&\frac{5 m \tilde{L}\left(\tilde{L}^4-2 \tilde{L}^2 L^2 \lambda_7-3 L^4 \mu_7\right)}{2  \ell_p^5 }
 \nonumber\\&=&\frac{5 m \tilde{L}^5  \left(1-f_{\infty}\lambda_7 \right)^2 }{2  \ell_p^5}
\end{eqnarray}
Where we have used the constraint condition
$\mu_7=-{\lambda^2\over 3}$ and $\tilde{L}^2=\frac{L^2}{f_\infty}$
in the last step. As a consistent check, if one turns off the
Lovelock gravity, one can reproduce the energy density of CFT which
is dual to pure Einstein gravity.

\end{document}